\documentclass[11pt,fleqn]{article}
\usepackage{amsmath,amssymb,amsthm} 

\newcommand{\p}{{\partial}}
\newcommand{\rank}{\mathop{\rm rank}\nolimits}
\newcommand{\Inv}{\mathop{\rm Inv}\nolimits}
\newcommand{\Int}{\mathop{\rm Int}\nolimits}

\newcommand{\diag}{\mathop{\rm diag}\nolimits}

\newtheorem{theorem}{Theorem}

\newtheorem{corollary}{Corollary}
\newtheorem{lemma}{Lemma}
{\theoremstyle{definition}

\newtheorem{note}{Note}
\newtheorem*{note*}{Note}
\newtheorem{proposition}{Proposition}
}

\begin{document}

\allowdisplaybreaks

\begin{flushleft}
\LARGE \bf Invariants of Triangular Lie Algebras
\end{flushleft}

\begin{flushleft} \bf
Vyacheslav Boyko~$^\dag$, Jiri Patera~$^\ddag$ and Roman Popovych~$^{\dag\S}$
\end{flushleft}

\noindent $^\dag$~Institute of Mathematics of NAS of Ukraine, 3
Tereshchenkivs'ka Str., Kyiv-4, 01601 Ukraine\\
$\phantom{^\dag}$~E-mail: boyko@imath.kiev.ua, rop@imath.kiev.ua

\noindent
$^\ddag$~Centre de Recherches Math\'ematiques,
Universit\'e de Montr\'eal,\\
$\phantom{^\ddag}$~C.P. 6128 succursale Centre-ville, Montr\'eal (Qu\'ebec), H3C 3J7 Canada\\
$\phantom{^\ddag}$~E-mail: patera@CRM.UMontreal.CA

\noindent $^\S$~Fakult\"at f\"ur Mathematik, Universit\"at Wien, Nordbergstra{\ss}e 15, A-1090 Wien, Austria

\begin{abstract}
\noindent
Triangular Lie algebras are the Lie algebras which can be faithfully represented
by triangular matrices of any finite size over the real/complex number field.
In the paper invariants (`generalized Casimir operators') are found for three classes of Lie algebras,
namely those which are either strictly or non-strictly triangular, and for so-called special upper triangular Lie algebras.
Algebraic algorithm of
[{\it J.\,Phys.\,A: Math.\,Gen.}, 2006, V.39, 5749; math-ph/0602046],
developed further in
[{\it J.\,Phys.\,A: Math.\,Theor.}, 2007, V.40, 113; math-ph/0606045], is used to determine the invariants.  A conjecture of
[{\it J.\,Phys.\,A: Math.\,Gen.}, 2001, V.34, 9085],
concerning the number of independent invariants and their form, is corroborated.
\end{abstract}

\section{Introduction}

The invariants of Lie algebras are one of their defining
characteristics. They have numerous applications in different fields
of mathematics and physics, in which Lie algebras arise
(representation theory, integrability of Hamiltonian differential
equations, quantum numbers etc). In particular, the polynomial
invariants of a Lie algebra exhaust its set of Casimir operators,
i.e., the center of its universal enveloping algebra. That is why
non-polynomial invariants are also called generalized Casimir
operators, and the usual Casimir operators are seen as `trivial'
generalized Casimir operators. Since the structure of invariants
strongly depends on the structure of the algebra and the
classification of all (finite-dimensional) Lie algebras is an
inherently difficult problem (actually unsolvable), it
seems to be impossible to elaborate a complete theory for
generalized Casimir operators in the general case. Moreover, if the
classification of a class of Lie algebras is known, then the
invariants of such algebras can be described exhaustively. These
problems have already been solved for the semi-simple and
low-dimensional Lie algebras, and also for the physically relevant
Lie algebras of fixed dimensions (see, e.g.,  references
in~\cite{Boyko&Patera&Popovych2006,Campoamor-Stursberg2007,Chaichian&Demichev&Nelipa1983,Patera&Sharp&Winternitz1976,Perroud1983}).

The actual problem is the investigation of generalized Casimir
operators for classes of solvable Lie algebras or non-solvable Lie
algebras with non-trivial radicals of arbitrary finite dimension.
There are a number of papers on the partial classification of such
algebras and the subsequent calculation of their invariants
\cite{Ancochea&Campoamor-Stursberg&GarciaVergnolle2006,
Campoamor-Stursberg2006a, Campoamor-Stursberg2007, Ndogmo2004,
Ndogmo&Wintenitz1994a, Ndogmo&Wintenitz1994b, Rubin&Winternitz1993,
Snobl&Winternitz2005, Tremblay&Winternitz1998,
Tremblay&Winternitz2001}.
In particular, Tremblay and Winternitz~\cite{Tremblay&Winternitz1998}
classified all the solvable Lie
algebras with the nilradicals isomorphic to the nilpotent algebra
$\mathfrak t_0(n)$ of strictly upper triangular matrices for any
fixed dimension $n$. Then in \cite{Tremblay&Winternitz2001}
invariants of these algebras were considered. The case $n=4$ was
investigated exhaustively. After calculating the invariants for a
sufficiently large value of~$n$, Tremblay and Winternitz made
conjectures for an arbitrary~$n$ on the number and form of
functionally independent invariants of the algebra $\mathfrak
t_0(n)$, and the `diagonal' solvable Lie algebras having $\mathfrak
t_0(n)$ as their nilradicals and possessing either the maximal
(equal to $n-1$) or minimal (one) number of nilindependent
elements. A statement on a functional basis of invariants was only
proved completely for the algebra $\mathfrak t_0(n)$. The
infinitesimal invariant criterion was used for the construction of
the invariants. Such an approach entails the necessity of solving a
system of $\rho$ first-order linear partial differential equations,
where $\rho$ has the order of the algebra's dimension. This is why
the calculations were very cumbersome and results were obtained due
to the thorough mastery of the method.

In this paper, we use our original algebraic method for the
construction of the invariants (`generalized Casimir operators') of
Lie algebras via the moving frames approach~\cite{Boyko&Patera&Popovych2006, Boyko&Patera&Popovych2007}.
The algorithm makes use of the knowledge of the associated inner
automorphism groups and Cartan's method of moving frames in its
Fels--Olver version~\cite{Fels&Olver1998,Fels&Olver1999}.
(For modern developments about the moving frame method and more
references, see also \cite{Olver&Pohjanpelto2007}.) Unlike standard
infinitesimal methods, it allows us to avoid solving systems of
differential equations, replacing them instead by algebraic
equations. As a result, the application of the algorithm is simpler.
Note that a closed approach was earlier proposed in~\cite{Kaneta1984a,Kaneta1984b,Perroud1983} for the specific case of inhomogeneous algebras.

The invariants of three classes of triangular Lie algebras are exhaustively investigated
(below $n$ is an arbitrary integer):
\begin{itemize}\itemsep=0ex
\item
nilpotent Lie algebras $\mathfrak t_0(n)$ of $n\times n$ strictly upper triangular matrices
(Section~\ref{SectionNilpotentAlgebraOfStrictlyUpperTriangleMatrices});
\item
solvable Lie algebras $\mathfrak t(n)$ of $n\times n$ upper triangular matrices
(Section~\ref{SectionSolvableAlgebraOfUpperTriangleMatrices});
\item
solvable Lie algebras $\mathfrak{st}(n)$ of $n\times n$ special upper triangular matrices
(Section~\ref{SectionSolvableAlgebraOfSpecialUpperTriangleMatrices}).
\end{itemize}
The triangular algebras are especially interesting due to their
`universality' properties. More precisely, any finite-dimensional
nilpotent Lie algebra is isomorphic to a subalgebra of $\mathfrak
t_0(n)$. Similarly, any finite-dimensional solvable Lie algebra
over an algebraically closed field of characteristic~0 (e.g., over~$\mathbb C$)
can be embedded as a subalgebra in $\mathfrak t(n)$ (or  $\mathfrak{st}(n)$).

We have adapted and optimized our algorithm for the specific case of
triangular Lie algebras via special double enumeration of basis
elements, individual choice of coordinates in the corresponding
inner automorphism groups and an appropriate modification of the
normalization procedure of the moving frame method. As a result, the
problems related to the construction of functional bases of
invariants are reduced for the algebras $\mathfrak t_0(n)$ and
$\mathfrak t(n)$ to solving \emph{linear systems of algebraic
equations}! Let us note that due to the natural embedding of
$\mathfrak{st}(n)$ to $\mathfrak t(n)$ and the representation
$\mathfrak t(n)=\mathfrak{st}(n)\oplus Z(\mathfrak t(n))$, where
$Z(\mathfrak t(n))$ is the center of $\mathfrak t(n)$, we can
construct a basis in the set of invariants of $\mathfrak{st}(n)$
without the usual calculations from a previously found basis in the
set of invariants of $\mathfrak t(n)$.

We re-prove the statement for a basis of invariants of $\mathfrak
t_0(n)$, which was first constructed
in~\cite{Tremblay&Winternitz2001} using the infinitesimal method in
a heuristic way, thereafter constructed
in~\cite{Boyko&Patera&Popovych2007} using an empiric technique based
on the exclusion of parameters within the framework of the algebraic
method. The aim of this paper in considering $\mathfrak t_0(n)$ is
to test and better understand the technique of working with
triangular algebras. The calculations for~$\mathfrak t(n)$ are
similar, albeit more complex, although they are much clearer and
easier than under the standard infinitesimal approach.

As proved in~\cite{Tremblay&Winternitz1998}, there is a unique
algebra with the nilradical $\mathfrak t_0(n)$ that contains a
maximum possible number ($n-1$) of nilindependent elements. A
conjecture on the invariants of this algebra is formulated in
Proposition~1 of~\cite{Tremblay&Winternitz2001}. We show that this algebra is isomorphic to
$\mathfrak{st}(n)$. As a result, the conjecture by Tremblay and
Winternitz on its invariants is effectively proved.


\section{The algorithm}\label{SectionAlgorithm}

The applied algebraic algorithm was first proposed
in~\cite{Boyko&Patera&Popovych2006} and then developed
in~\cite{Boyko&Patera&Popovych2007}. Ibid it was effectively tested
for the low-dimensional Lie algebras and a wide range of solvable Lie algebras with a fixed structure of nilradicals.
 The presentation of the algorithm
here differs from~\cite{Boyko&Patera&Popovych2006,
Boyko&Patera&Popovych2007}, the differences being important within
the framework of applications.

For convenience of the reader and to introduce some necessary
notations, before the description of the algorithm, we briefly
repeat the preliminaries given
in~\cite{Boyko&Patera&Popovych2006,Boyko&Patera&Popovych2007} about
the statement of the problem of calculating Lie algebra invariants,
and on the implementation of the moving frame
method~\cite{Fels&Olver1998,Fels&Olver1999}. The comparative
analysis of the standard infinitesimal and the presented algebraic
methods, as well as their modifications, is given in the second part
of this section.

Consider a Lie algebra~$\mathfrak g$ of dimension $\dim \mathfrak
g=n<\infty$ over the complex or real field and the corresponding
connected Lie group~$G$. Let~$\mathfrak g^*$ be the dual space of
the vector space~$\mathfrak g$. The map ${\rm Ad}^*\colon G\to
GL(\mathfrak g^*)$, defined for any $g\in G$ by the relation
\[
\langle{\rm Ad}^*_g x,u\rangle=\langle x,{\rm Ad}_{g^{-1}}u\rangle
\quad \mbox{for all $x\in \mathfrak g^*$ and $u\in \mathfrak g$}
\]
is called the {\it coadjoint representation} of the Lie group~$G$.
Here ${\rm Ad}\colon G\to GL(\mathfrak g)$ is the usual adjoint
representation of~$G$ in~$\mathfrak g$, and the image~${\rm Ad}_G$
of~$G$ under~${\rm Ad}$ is the inner automorphism group ${\rm
Int}(\mathfrak g)$ of the Lie algebra~$\mathfrak g$. The image
of~$G$ under~${\rm Ad}^*$ is a subgroup of~$GL(\mathfrak g^*)$ and
is denoted by~${\rm Ad}^*_G$.

A function $F\in C^\infty(\mathfrak g^*)$ is called an {\it
invariant} of~${\rm Ad}^*_G$ if $F({\rm Ad}_g^* x)=F(x)$ for all
$g\in G$ and $x\in \mathfrak g^*$. The set of invariants of ${\rm
Ad}^*_G$ is denoted by $\Inv({\rm Ad}^*_G)$. The maximal number
$N_\mathfrak g$ of functionally independent invariants in $\Inv({\rm
Ad}^*_G)$ coincides with the codimension of the regular orbits
of~${\rm Ad}^*_G$, i.e., it is given by the difference
\[
N_\mathfrak g=\dim \mathfrak g-\rank {\rm Ad}^*_G.
\]
Here $\rank {\rm Ad}^*_G$ denotes the dimension of the regular
orbits of~${\rm Ad}^*_G$ and will be called the {\em rank of the
coadjoint representation} of~$G$ (and of~$\mathfrak g$). It is a
basis independent characteristic of the algebra~$\mathfrak g$, the
same as $\dim \mathfrak g$ and $N_\mathfrak g$.

To calculate the invariants explicitly, one should fix a basis
$\mathcal E=\{e_1,\ldots,e_n\}$ of the algebra~$\mathfrak g$. It
leads to fixing the dual basis $\mathcal E^*=\{e_1^*,\ldots,e_n^*\}$
in the dual space~$\mathfrak g^*$ and to the identification of ${\rm
Int}(\mathfrak g)$ and ${\rm Ad}^*_G$ with the associated matrix
groups. The basis elements $e_1,\ldots,e_n$ satisfy the commutation
relations $[e_i,e_j]=\sum_{k=1}^{n}c_{ij}^k e_k$, $i,j=1,\ldots,n$,
where $c_{ij}^k$ are components of the tensor of structure constants
of~$\mathfrak g$ in the basis~$\mathcal E$.

Let $x\to\check x=(x_1,\ldots,x_n)$ be the coordinates in~$\mathfrak
g^*$ associated with~$\mathcal E^*$. Given any invariant
$F(x_1,\ldots,x_n)$ of~${\rm Ad}^*_G$, one finds the corresponding
invariant of the Lie algebra~$\mathfrak g$ by symmetrization,
$\mathop{\rm Sym}\nolimits F(e_1,\ldots,e_n)$, of $F$. It is often
called a \emph{generalized Casimir operator} of~$\mathfrak g$. If
$F$ is a~polynomial, $\mathop{\rm Sym}\nolimits F(e_1,\ldots,e_n)$
is a usual \emph{Casimir operator}, i.e., an element of the center of
the universal enveloping algebra of~$\mathfrak g$. More precisely,
the symmetrization operator~$\mathop{\rm Sym}\nolimits$ acts only on
the monomials of the forms~$e_{i_1}\cdots e_{i_r}$, where there are
non-commuting elements among~$e_{i_1}, \ldots, e_{i_r}$, and is
defined by the formula
\[
\mathop{\rm Sym}\nolimits (e_{i_1}\cdots e_{i_r})=\dfrac1{r!}\sum_{\sigma\in S_r}e_{i_{\sigma_1}}\cdots e_{i_{\sigma_r}},
\]
where $i_1, \ldots, i_r$ take values from 1 to $n$, $r\geqslant 2$.
The symbol $S_r$ denotes the permutation group consisting of $r$
elements. The set of invariants of $\mathfrak g$ is denoted by
$\Inv(\mathfrak g)$.

A set of functionally independent invariants $F^l(x_1,\ldots,x_n)$,
\mbox{$l=1,\ldots,N_\mathfrak g$}, forms {\it a~functional basis}
({\it fundamental invariant}) of $\Inv({\rm Ad}^*_G)$, i.e., any
invariant $F(x_1,\ldots,x_n)$ can be uniquely represented as
a~function of~$F^l(x_1,\ldots,x_n)$, \mbox{$l=1,\ldots,N_\mathfrak
g$}. Accordingly the set of $\mathop{\rm Sym}\nolimits
F^l(e_1,\ldots,e_n)$, \mbox{$l=1,\ldots,N_\mathfrak g$}, is called a
basis of~$\Inv(\mathfrak g)$.

Our task here is to determine the basis of the functionally
independent invariants for ${\rm Ad}^*_G$, and then to transform
these invariants into the invariants of the algebra~$\mathfrak g$.
Any other invariant of $\mathfrak g$ is a function of the
independent ones.

Let us recall some facts from \cite{Fels&Olver1998,Fels&Olver1999}
and adapt them to the particular case of the coadjoint action of~$G$
on~$\mathfrak g^*$. Let~$\mathcal{G}={\rm Ad}^*_G\times \mathfrak
g^*$ denote the trivial left principal ${\rm Ad}^*_G$-bundle
over~$\mathfrak g^*$. The right regularization~$\widehat R$ of the
coadjoint action of~$G$ on~$\mathfrak g^*$ is the diagonal action
of~${\rm Ad}^*_G$ on~$\mathcal{G}={\rm Ad}^*_G\times \mathfrak g^*$.
It is provided by the map
$
\widehat R_g({\rm Ad}^*_h,x)=({\rm Ad}^*_h\cdot {\rm Ad}^*_{g^{-1}},{\rm Ad}^*_g x),
\ g,h\in G, \ x\in \mathfrak g^*,
$
where the action on the bundle~$\mathcal{G}={\rm Ad}^*_G\times
\mathfrak g^*$ is regular and free. We call \raisebox{0ex}[0ex][0ex]{$\widehat R_g$} the
\emph{lifted coadjoint action} of~$G$. It projects back to the
coadjoint action on~$\mathfrak g^*$ via the ${\rm
Ad}^*_G$-equivariant projection~$\pi_{\mathfrak g^*}\colon
\mathcal{G}\to \mathfrak g^*$. Any \emph{lifted invariant} of~${\rm
Ad}^*_G$ is a (locally defined) smooth function from~$\mathcal{G}$
to a~manifold, which is invariant with respect to the lifted
coadjoint action of~$G$. The function $\mathcal
I\colon\mathcal{G}\to \mathfrak g^*$ given by $\mathcal I=\mathcal
I({\rm Ad}^*_g,x)={\rm Ad}^*_g x$ is the \emph{fundamental lifted
invariant} of ${\rm Ad}^*_G$, i.e., $\mathcal I$ is a lifted
invariant, and any lifted invariant can be locally written as a
function of~$\mathcal I$. Using an arbitrary function~$F(x)$
on~$\mathfrak g^*$, we can produce the lifted
invariant~$F\circ\mathcal I$ of~${\rm Ad}^*_G$ by replacing $x$ with
$\mathcal I={\rm Ad}^*_g x$ in the expression for~$F$. Ordinary
invariants are particular cases of lifted invariants, where one
identifies any invariant formed as its composition with the standard
projection~$\pi_{\mathfrak g^*}$. Therefore, ordinary invariants are
particular functional combinations of lifted ones that happen to be
independent of the group parameters of~${\rm Ad}^*_G$.

The \emph{algebraic algorithm} for finding invariants of the Lie
algebra $\mathfrak g$ is briefly formulated in the following four
steps.

\medskip

1. {\it Construction of the generic matrix $B(\theta)$ of~${\rm
Ad}^*_G$.} $B(\theta)$ is the matrix of an inner automorphism of the
Lie algebra~$\mathfrak g$ in the given basis $e_1$, \ldots, $e_n$,
$\theta=(\theta_1,\ldots,\theta_r)$ is a complete tuple of group
parameters (coordinates) of~$\mathop{\rm Int}(\mathfrak g)$, and
$r=\dim{\rm Ad}^*_G=\dim\mathop{\rm Int}(\mathfrak g)=n-\dim{\rm
Z}(\mathfrak g),$ where ${\rm Z}(\mathfrak g)$ is the center
of~$\mathfrak g$.

\medskip

2. {\it Representation of the fundamental lifted invariant.} The
explicit form of the fundamental lifted invariant~$\mathcal
I=(\mathcal I_1,\ldots,\mathcal I_n)$ of ${\rm Ad}^*_G$ in the
chosen coordinates~$(\theta,\check x)$ in ${\rm Ad}^*_G\times
\mathfrak g^*$ is $\mathcal I=\check x\cdot B(\theta)$, i.e.,
$
(\mathcal I_1,\ldots,\mathcal I_n)=(x_1,\ldots,x_n)\cdot B(\theta_1,\ldots,\theta_r).
$

\medskip

3. {\it Elimination of parameters by normalization}. We choose the
maximum possible number $\rho$ of lifted invariants $\mathcal
I_{j_1}$, \ldots, $\mathcal I_{j_\rho}$, constants $c_1$, \ldots,
$c_\rho$ and group parameters
$\theta_{k_1}$,~\ldots,~$\theta_{k_\rho}$ such that the equations
$\mathcal I_{j_1}=c_1$, \ldots, $\mathcal I_{j_\rho}=c_\rho$ are
solvable with respect to $\theta_{k_1}$,~\ldots,~$\theta_{k_\rho}$.
After substituting the found values of
$\theta_{k_1}$,~\ldots,~$\theta_{k_\rho}$ into the other lifted
invariants, we obtain $N_\mathfrak g=n-\rho$ expressions $F^l
(x_1,\ldots,x_n)$ without $\theta$'s.

\medskip

4. {\it Symmetrization.} The functions $F^l(x_1,\ldots,x_n)$
necessarily form a basis of~$\Inv({\rm Ad}^*_G)$. They are
symmetrized to $\mathop{\rm Sym}\nolimits F^l(e_1,\ldots,e_n)$. It
is the desired basis of~$\Inv(\mathfrak g)$.

\medskip

Let us give some remarks on the steps of the algorithm, mainly
paying attention to the special features of its variation in this
paper, and where it differs from the conventional infinitesimal
method.

Usually, the second canonical coordinate on $\Int(\mathfrak g)$ is
enough for the first step, although sometimes, the first canonical
coordinate on $\Int(\mathfrak g)$ is the more appropriate choice. In
both the cases, the matrix $B(\theta)$ is calculated by exponentiation
from matrices associated with the structure constants. Often the
parameters~$\theta$ are additionally transformed in a trivial manner
(signs, renumbering, re-denotation etc) for simplification of the
final presentation of~$B(\theta)$. It is also sometimes convenient
for us to introduce `virtual' group parameters corresponding to the
center basis elements. Efficient exploitation of the algorithm
imposes certain constrains on the choice of bases for $\mathfrak g$,
in particular, in the enumeration of their elements; thus
automatically yielding simpler expressions for elements
of~$B(\theta)$ and, therefore, expressions of the lifted invariants.
In some cases the simplification is considerable.

In contrast with the general situation, for the triangular Lie
algebras we use special coordinates for their inner automorphism
groups, which naturally harmonize with the canonical matrix
representations of the corresponding Lie groups and with special
`matrix' enumeration of the basis elements. The application of the
individual approach results in the clarification and a substantial
reduction of all calculations. In particular, algebraic systems
solved under normalization become linear with respect to their
parameters.

Since $B(\theta)$ is a general form matrix from $\Int(\mathfrak g)$,
it should not be adapted in any way for the second step.

Indeed, the third step of the algorithm can involve different
techniques of elimination of parameters which are also based on
using an explicit form of lifted invariants
\cite{Boyko&Patera&Popovych2006,Boyko&Patera&Popovych2007}. The
applied normalization procedure~\cite{Fels&Olver1998,Fels&Olver1999}
can also be subject to some variations and can applied in a more
involved manner.

As a rule, in complicated cases the main difficulty is created
by the determination of the number~$\rho$, who is actually
equal to $\rank {\rm Ad}^*_G$, which is equivalent to finding the
maximum number $N_\mathfrak g$ of functionally independent
invariants in $\Inv({\rm Ad}^*_G)$, since $N_\mathfrak g=\dim
\mathfrak g-\rank {\rm Ad}^*_G$. The rank $\rho$ of the coadjoint
representation ${\rm Ad}^*_G$ can be calculated in different ways,
e.g., by the closed formulas
\[
\rho=\max_{\check x\in\mathbb{R}^n}\rank\, \biggl(\sum_{k=1}^{n}c_{ij}^k x_k\biggr)_{i,j=1}^n, \qquad
\rho=\max_{\check x\in\mathbb{R}^n}\max_{\theta\in\mathbb{R}^r}\rank \dfrac{\p\mathcal I}{\p\theta}
\]
or with the use of indirect argumentation. The first formula is
native to the infinitesimal approach to invariants (see, e.g.,
\cite{Campoamor-Stursberg2004, Ndogmo&Wintenitz1994b,
Patera&Sharp&Winternitz1976, Tremblay&Winternitz2001} and other
references) since it gives the number of algebraically independent
differential equations in the linear system of first-order partial
differential equations $\sum_{j,k=1}^{n}c_{ij}^k x_kF_{x_j}=0$,
which arises under this approach and is the infinitesimal criterion
for invariants of the algebra~$\mathfrak g$ under the fixed
basis~$\mathcal E$. The second formula shows that $\rank {\rm
Ad}^*_G$ coincides with the maximum dimension of a nonsingular
submatrix in the Jacobian matrix~$\p\mathcal I/\p\theta$. The tuples
of lifted invariants and parameters associated with this submatrix
are appropriate for the normalization procedure, where the
constants $c_1$, \ldots, $c_\rho$ are chosen to lie in the range of
values of the corresponding lifted invariants.

If $\rho$ is known then the sufficient number ($N_\mathfrak g=\dim
\mathfrak g-\rho$) of functionally independent invariants can be
found with various `empiric' techniques in the frameworks of both
the infinitesimal and algebraic approaches. For example, expressions
of candidates for invariants can be deduced from invariants of
similar low-dimensional Lie algebras and then tested via
substitution to the infinitesimal criterion for invariants. It is
the method used in \cite{Tremblay&Winternitz2001} to describe
invariants of the Lie algebra~$\mathfrak t_0(n)$ of strictly upper
triangular $n\times n$ matrices for any fixed~$n\geqslant2$. In the
framework of the algebraic approach, invariants can be constructed
via the combination of lifted invariants in expressions not
depending on the group parameters~\cite{Fels&Olver1998,Fels&Olver1999}.
This method was applied, in particular, to
low-dimensional algebras and the algebra~$\mathfrak t_0(n)$
\cite{Boyko&Patera&Popovych2006,Boyko&Patera&Popovych2007}.
Other empiric techniques, e.g.,
based on commutator properties~\cite{Barannyk&Fushchych1986} also can be used.

At the same time, a basis of $\Inv({\rm Ad}^*_G)$ may be constructed
without first determining the number of basis elements. Since under
such consideration the infinitesimal approach leads to the necessity
of the complete integration of the partial differential equations
from the infinitesimal invariant criterion, the domain of its
applicability seems quite narrow (low-dimensional algebras and Lie
algebra of special simple structure). A similar variation of the
algebraic method is based on the following obvious statement.

\begin{proposition}\label{PropositionOnNormalization}
Let~$\mathcal I=(\mathcal I_1,\ldots,\mathcal I_n)$ be a fundamental
lifted invariant, for the lifted invariants $\mathcal I_{j_1}$,
\ldots, $\mathcal I_{j_\rho}$ and some constants $c_1$, \ldots,
$c_\rho$ the system $\mathcal I_{j_1}=c_1$, \ldots, $\mathcal
I_{j_\rho}=c_\rho$ be solvable with respect to the parameters
$\theta_{k_1}$,~\ldots,~$\theta_{k_\rho}$ and substitution of the
found values of $\theta_{k_1}$,~\ldots,~$\theta_{k_\rho}$ into the
other lifted invariants result in $m=n-\rho$ expressions
$\hat{\mathcal I}_l$, $l=1,\dots,m$, depending only on $x$'s. Then
$\rho=\rank {\rm Ad}^*_G$, $m=N_\mathfrak g$ and $\hat{\mathcal
I}_1$, \ldots, $\hat{\mathcal I}_m$ form a basis of $\Inv({\rm
Ad}^*_G)$.
\end{proposition}

Our experience on the calculation of invariants of a wide range of
Lie algebras shows that the version of the algebraic method, which
is based on Proposition~\ref{PropositionOnNormalization}, is most
effective. It is the version that is used in this paper.

Note that the normalization procedure is difficult to be made
algorithmic. There is a big ambiguity in the choice of the
normalization equations. We can take different tuples of $\rho$
lifted invariants and $\rho$ constants, which lead to systems
solvable with respect to $\rho$ parameters. Moreover, lifted
invariants can be additionally combined before forming a system of
normalization equations or substitution of found values of
parameters. Another possibility is to use a floating system of
normalization equations (see Section~6.2
of~\cite{Boyko&Patera&Popovych2007}). This means that elements of an
invariant basis are constructed under different normalization
constraints. The choice of an optimal method results in a
considerable reduction of calculations and a practical form of
constructed invariants.


\section{Nilpotent algebra of strictly upper triangular matrices}\label{SectionNilpotentAlgebraOfStrictlyUpperTriangleMatrices}

Consider the nilpotent Lie algebra $\mathfrak t_0(n)$ isomorphic to
the one of the strictly upper triangular $n\times n$ matrices over the
field $\mathbb F$, where $\mathbb F$ is either $\mathbb C$ or
$\mathbb R$. $\mathfrak t_0(n)$ has dimension $n(n-1)/2$. It is the
Lie algebra of the Lie group ${T}_0(n)$ of upper unipotent $n\times
n$ matrices, i.e., upper triangular matrices with entries equal to 1 in the diagonal.

As mentioned above, the basis of $\Inv(\mathfrak t_0(n))$ was first
constructed in a heuristic way in~\cite{Tremblay&Winternitz2001}
within the framework of the infinitesimal approach. This result was
re-obtained in~\cite{Boyko&Patera&Popovych2007} with the use of the
pure algebraic algorithm first proposed
in~\cite{Boyko&Patera&Popovych2006} and developed
in~\cite{Boyko&Patera&Popovych2007}. Also, it is the unique example included among the
wide variety of solvable Lie algebras investigated
in~\cite{Boyko&Patera&Popovych2007}, in which the `empiric'
technique of excluding group parameters from lifted invariants was
applied. Although this technique was very effective in constructing
a set of functionally independent invariants (calculations were
reduced via a special representation of the coadjoint action to a
trivial identity using matrix determinants, see
Note~\ref{NoteOnTrickWithStrictlyUpperTriangularMatrices}), the main
difficulty was in proving that the set of invariants is a basis of
$\Inv(\mathfrak t_0(n))$, i.e.\ cardinality of the set equals the
maximum possible number of functionally independent invariants.
Under the infinitesimal approach~\cite{Tremblay&Winternitz2001} the
main difficulty was the same.

In this section we construct a basis of $\Inv(\mathfrak t_0(n))$
with the algebraic algorithm but exclude group parameters from
lifted invariants by the normalization procedure. In contrast with
the previous expositions (Section 3
of~\cite{Tremblay&Winternitz2001} and Section 8
of~\cite{Boyko&Patera&Popovych2007}), sufficiency of the number of
found invariants for forming a basis of $\Inv(\mathfrak t_0(n))$ is
proved in the process of calculating them. Investigation of
$\Inv(\mathfrak t_0(n))$ in this way gives us a sense of the
specific features of the normalization procedure in the case of Lie
algebras having nilradicals isomorphic (or closed) to $\mathfrak
t_0(n)$.

For the algebra $\mathfrak t_0(n)$ we use a `matrix' enumeration of
the basis elements with an `increasing' pair of indices, in a
similar way to the canonical basis $\{E^n_{ij},\,i<j\}$ of the
isomorphic matrix algebra.

Hereafter $E^n_{ij}$ (for fixed values of $i$ and $j$) denotes the
$n\times n$ matrix $(\delta_{ii'}\delta_{jj'})$ with $i'$ and $j'$
running the numbers of rows and column respectively, i.e., the
$n\times n$ matrix with a unit element on the cross of the $i$-th
row and the $j$-th column, and zero otherwise.
$E^n=\diag(1,\ldots,1)$ is the $n\times n$ unity matrix. The indices
$i$, $j$, $k$ and $l$ run at most from 1 to~$n$. Only additional
constraints on the indices are indicated.

Thus, the basis elements $e_{ij}\sim E^n_{ij}$, $i<j$, satisfy the commutation relations
\[
[e_{ij},e_{i'\!j'\!}]=\delta_{i'\!j}e_{ij'\!}-\delta_{ij'}e_{i'\!j},
\]
where $\delta_{ij}$ is the Kronecker delta.

Let $e_{ji}^*$, $x_{ji}$ and $y_{ij}$ denote the basis element and
the coordinate function in the dual space $\mathfrak t_0^*(n)$ and
the coordinate function in~$\mathfrak t_0(n)$, which correspond to
the basis element~$e_{ij}$, $i<j$.
In~particular, $\langle e_{j'\!i'\!}^*,e_{ij}\rangle=\delta_{ii'\!}\delta_{jj'\!}.$
The reverse order of subscripts of the objects
associated with the dual space~$\mathfrak t_0^*(n)$
is justified by the simplification of a matrix representation of lifted invariants.
We~complete the sets of $x_{ji}$
and $y_{ij}$ in the matrices $X$ and $Y$ with zeros. Hence $X$ is a
strictly lower triangular matrix and $Y$ is a strictly upper triangular
one.

We reproduce Lemma~1 from~\cite{Boyko&Patera&Popovych2007} together
with its proof, since it is important for further consideration.

\begin{lemma}\label{LemmaOnLiftedInvsOfStrictlyUpperTriangularMatrices}
A complete set of independent lifted invariants of ${\rm Ad}^*_{T_0(n)}$
is exhaustively given by the expressions
\[
\mathcal I_{ij}=x_{ij}+\sum_{i<i'}b_{ii'}x_{i'\!j}+\sum_{j'<j}b_{j'\!j}x_{ij'}
+\sum_{i<i'\!,\,j'<j}b_{ii'}\widehat b_{j'\!j}x_{i'\!j'}, \quad j<i,
\]
where $B=(b_{ij})$ is an arbitrary matrix from ${T}_0(n)$, and
$B^{-1}=(\widehat b_{ij})$ is the inverse matrix of~$B$.
\end{lemma}

\begin{proof} The adjoint action of $B\in{T}_0(n)$ on the matrix~$Y$ is
${\rm Ad}_BY=BYB^{-1}$, i.e.,
\[
{\rm Ad}_B\sum_{i<j}y_{ij}e_{ij}=\sum_{i<j}(BYB^{-1})_{ij}e_{ij}
=\sum_{i\leqslant i'<j'\leqslant j}b_{ii'}y_{i'\!j'}\widehat b_{j'\!j}e_{ij}.
\]
After changing $e_{ij}\to x_{ji}$, $y_{ij}\to e_{ji}^*$,
$b_{ij}\leftrightarrow \widehat b_{ij}$ in the latter equality, we
obtain the representation of the coadjoint action of~$B$
\[
{\rm Ad}_B^*\sum_{i<j}x_{ji}e_{ji}^*
=\sum_{i\leqslant i'<j'\leqslant j}b_{j'\!j}x_{ji}\widehat b_{ii'}e_{j'\!i'}^*
=\sum_{i'<j'}(BXB^{-1})_{j'\!i'}e_{j'\!i'}^*.
\]
Therefore, the elements $\mathcal I_{ij}$, $j<i$, of the matrix
$
\mathcal I=BXB^{-1}, \ B\in{T}_0(n),
$
form a complete set of the independent lifted invariants of ${\rm
Ad}^*_{T_0(n)}$.
\end{proof}

\begin{note}
The center of the group ${T}_0(n)$ is
$Z({T}_0(n))=\{E^n+b_{1n}E^n_{1n},\ b_{1n}\in\mathbb F\}$. The inner
automorphism group of~$\mathfrak t_0(n)$ is isomorphic to the
factor-group ${T}_0(n)/Z({T}_0(n))$ and hence its dimension is
$\frac12n(n-1)-1$. The parameter $b_{1n}$ in the above
representation of the lifted invariants is not essential.
\end{note}

Below $A^{i_1,i_2}_{j_1,j_2}$, where $i_1\leqslant i_2$,
$j_1\leqslant j_2$, denotes the submatrix
$(a_{ij})^{i=i_1,\ldots,i_2}_{j=j_1,\ldots,j_2}$ of a matrix
$A=(a_{ij})$. The conjugate value of $k$ with respect to $n$ is
denoted by $\varkappa$, i.e. $\varkappa=n-k+1$. The standard
notation $|A|=\det A$ is used.

\begin{theorem}\label{TheoremOnBasisOfInvsOfCoadjRepresentationOfStrictlyUpperTriangularMatrices}
A basis of $\Inv({\rm Ad}^*_{T_0(n)})$ consists of the polynomials
\[
|X^{\varkappa,n}_{1,k}|, \quad k=1, \ldots, \left[\frac n2\right].
\]
\end{theorem}

\begin{proof}
Under normalization we impose the following restriction on the
lifted invariants $\mathcal I_{ij}$, $j<i$:
\[
\mathcal I_{ij}=0 \quad\mbox{if}\quad j<i,\ (i,j)\not=(n-j'+1,j'),\ j'=1,\ldots,\left[\frac{n}2\right].
\]
It means that we do not only fix the values of the elements of the
lifted invariant matrix~$\mathcal I$, which are situated on the
secondary diagonal, under the main diagonal. The other significant
elements of~$\mathcal I$ are given the value 0.
As shown below, the chosen normalization is correct since it provides satisfying the conditions of
Proposition~\ref{PropositionOnNormalization}.

In view of the (triangular) structure of the matrices $B$ and $X$ the
formula $\mathcal I=BXB^{-1}$, determining the lifted invariants
implies that $BX=\mathcal IB$. This matrix equality is also
significant for the matrix elements underlying the main diagonals of
the left and right hand sides, i.e.,
\[
x_{ij}+\sum_{i<i'}b_{ii'}x_{i'\!j}=\mathcal I_{ij}+\sum_{j'<j}\mathcal I_{ij'}b_{j'\!j}, \quad j<i.
\]
For convenience we divide the latter system under the chosen
normalization conditions into four sets of subsystems
\begin{gather*}
S_1^k\colon\qquad x_{\varkappa j}+\sum_{i'>\varkappa}b_{\varkappa i'}x_{i'\!j}=0, \qquad
i=\varkappa,\quad j<k,\quad k=2,\ldots,\left[\frac n2\right],
\\
S_2^k\colon\qquad x_{\varkappa k}+\sum_{i'>\varkappa}b_{\varkappa i'}x_{i'\!k}=\mathcal I_{\varkappa k}, \qquad
i=\varkappa,\quad j=k,\quad k=1,\ldots,\left[\frac n2\right],
\\
S_3^k\colon\qquad x_{\varkappa j}+\sum_{i'>\varkappa}b_{\varkappa i'}x_{i'\!j}=\mathcal I_{\varkappa k}b_{kj}, \qquad
i=\varkappa,\quad k<j<\varkappa,\quad k=1,\ldots,\left[\frac n2\right]-1,
\\
S_4^k\colon\qquad x_{kj}+\sum_{i'>k}b_{ki'}x_{i'\!j}=0, \qquad
i=k,\quad j<k,\quad k=2,\ldots,\left[\frac{n+1}2\right],
\end{gather*}
and solve them one by one. The subsystem~$S_2^1$ consists of the
single equation $\mathcal I_{n1}=x_{n1}$ which gives the simplest
form of the invariant corresponding to the center of the algebra
$\mathfrak t_0(n)$. For any fixed $k\in\{2,\dots,[n/2]\}$ the
subsystem $S_1^k \cup S_2^k$ is a well-defined system of linear
equations with respect to $b_{\varkappa i'}$, $i'>\varkappa$, and
$\mathcal I_{\varkappa k}$. Solving it, e.g., by the Cramer method,
we obtain that $b_{\varkappa i'}$, $i'>\varkappa$, are expressions
of $x_{i'\!j}$, $i'>\varkappa$, $j<k$, the explicit form of which is
not essential in what follows, and
\[
\mathcal I_{\varkappa k}=(-1)^{k+1}\frac{|X^{\varkappa,n}_{1,k}|}{|X^{\varkappa+1,n}_{1,k-1}|},
\quad k=2, \ldots, \left[\frac n2\right].
\]
The combination of the found values of $\mathcal I_{\varkappa k}$
results in the invariants from the statement of the theorem. The
functional independence of these invariants is obvious.

After substituting the expressions of $\mathcal I_{\varkappa k}$ and
$b_{\varkappa i'}$, $i'>\varkappa$, via $x$'s, into $S_3^k$, we
trivially resolve $S_3^k$ with respect to $b_{kj}$ as an uncoupled
system of linear equations. In performing the subsequent
substitution of the calculated expressions for $b_{kj}$ to $S_4^k$,
for any fixed~$k$, we obtain a well-defined system of linear
equations, e.g., with respect to $b_{ki'}$, $i'>\varkappa$.

Under the normalization we express the non-normalized lifted
invariants via $x$'s and find a part of the parameters $b$'s of the
coadjoint action via $x$'s and the other $b$'s.
No equations involving only $x$'s are obtained.
In view of Proposition~\ref{PropositionOnNormalization}, this implies
that the choice of the normalization constraints is correct and,
therefore, the number of functionally independent invariants found is maximal, i.e.,
they form a basis of $\Inv({\rm Ad}^*_{T_0(n)})$.
\end{proof}

\begin{corollary}\label{CorollaryOnBasisOfInvsOfStrictlyUpperTriangularMatrices}
A basis of~$\Inv(\mathfrak t_0(n))$ is formed by the Casimir
operators
\[
\det(e_{ij})^{i=1,\ldots,k}_{j=n-k+1,\ldots,n}, \quad k=1, \ldots, \left[\frac n2\right].
\]
\end{corollary}
\begin{proof}
Since the basis elements corresponding the coordinate functions from
the constructed basis of~$\Inv({\rm Ad}^*_{T_0(n)})$ commute, the
symmetrization procedure is trivial.
\end{proof}

\begin{note}\label{NoteOnTrickWithStrictlyUpperTriangularMatrices}
The set of the invariants from Theorem~\ref{TheoremOnBasisOfInvsOfCoadjRepresentationOfStrictlyUpperTriangularMatrices}
can be easily found from the equality \mbox{$\mathcal I=BXB^{-1}$} by the
following empiric trick used in Lemma~2 from~\cite{Boyko&Patera&Popovych2007}.
For any fixed $k\in\{1,\dots,[n/2]\}$ we restrict the equality to the submatrix
with the row range $\varkappa,\ldots,n$ and the column range $1,\ldots,k$:
$\mathcal{I}^{\varkappa,n}_{1,k}=B^{\varkappa,n}_{\varkappa,n}X^{\varkappa,n}_{1,k}(B^{-1})^{1,k}_{1,k}$.
Since $|B^{\varkappa,n}_{\varkappa,n}|=|(B^{-1})^{1,k}_{1,k}|=1$, we
obtain $|\mathcal{I}^{\varkappa,n}_{1,k}|=|X^{\varkappa,n}_{1,k}|$,
i.e., $|X^{\varkappa,n}_{1,k}|$ is an invariant of~${\rm Ad}^*_{T_0(n)}$ in view of the definition of an invariant.
Functional independence of the constructed invariants is obvious.
The proof of $N_{\mathfrak t_0(n)}=[n/2]$ is much more difficult (see Lemma~3~of~\cite{Boyko&Patera&Popovych2007}).
\end{note}


\section{Solvable algebra of upper triangular matrices}\label{SectionSolvableAlgebraOfUpperTriangleMatrices}

In a way analogous to the previous section, consider the solvable
Lie algebra $\mathfrak t(n)$ isomorphic to the one of upper triangular
$n\times n$ matrices. $\mathfrak t(n)$  has dimension $n(n+1)/2$. It
is the Lie algebra of the Lie group $T(n)$ of nonsingular upper
triangular $n\times n$ matrices.

Its basis elements are convenient to enumerate with a
`non-decreasing' pair of indices in a similar way to the canonical
basis $\{E^n_{ij},\,i\leqslant j\}$ of the isomorphic matrix
algebra. Thus, the basis elements $e_{ij}\sim E^n_{ij}$, $i\leqslant
j$, satisfy the commutation relations
\[
[e_{ij},e_{i'\!j'\!}]=\delta_{i'\!j}e_{ij'\!}-\delta_{ij'}e_{i'\!j},
\]
where $\delta_{ij}$ is the Kronecker delta.

Hereafter the indices $i$, $j$, $k$ and $l$ again run at most from 1 to~$n$.
Only additional constraints on the indices are indicated.

The center of $\mathfrak t(n)$ is one-dimensional and coincides with
the linear span of the sum $e_{11}+\dots+e_{nn}$ corresponding to
the unity matrix~$E^n$. The elements $e_{ij}$, $i<j$, and
$e_{11}+\dots+e_{nn}$ form a basis of the nilradical of  $\mathfrak
t(n)$, which is isomorphic to $\mathfrak t_0(n)\oplus \mathfrak a$.
Here $\mathfrak a$ is the one-dimensional (Abelian) Lie algebra.

Let $e_{ji}^*$, $x_{ji}$ and $y_{ij}$ denote the basis element and
the coordinate function in the dual space $\mathfrak t^*(n)$ and the
coordinate function in~$\mathfrak t(n)$, which correspond to the
basis element~$e_{ij}$, $i\leqslant j$.
Thus, $\langle e_{j'\!i'\!}^*,e_{ij}\rangle=\delta_{ii'\!}\delta_{jj'\!}.$
We complete the sets of
$x_{ji}$ and $y_{ij}$ in the matrices $X$ and $Y$ with zeros. Hence
$X$ is a lower triangular matrix and $Y$ is an upper triangular one.

\begin{lemma}\label{LemmaOnLiftedInvsOfUpperTriangularMatrices}
A fundamental lifted invariant of ${\rm Ad}^*_{T(n)}$ is formed by
the expressions
\[
\mathcal I_{ij}=\sum_{i\leqslant i'\!,\,j'\leqslant j}b_{ii'}\widehat b_{j'\!j}x_{i'\!j'}, \quad j\leqslant i,
\]
where $B=(b_{ij})$ is an arbitrary matrix from $T(n)$, and
$B^{-1}=(\widehat b_{ij})$ is the inverse matrix of~$B$.
\end{lemma}

\begin{proof} The adjoint action of $B\in T(n)$ on the matrix~$Y$ is ${\rm Ad}_BY=BYB^{-1}$, i.e.
\[
{\rm Ad}_B\sum_{i\leqslant j}y_{ij}e_{ij}=\sum_{i\leqslant j}(BYB^{-1})_{ij}e_{ij}
=\sum_{i\leqslant i'\leqslant j'\leqslant j}b_{ii'}y_{i'\!j'}\widehat b_{j'\!j}e_{ij}.
\]
After changing $e_{ij}\to x_{ji}$, $y_{ij}\to e_{ji}^*$,
$b_{ij}\leftrightarrow \widehat b_{ij}$ in the latter equality, we
obtain the representation for the coadjoint action of~$B$
\[
{\rm Ad}_B^*\sum_{i\leqslant j}x_{ji}e_{ji}^*
=\sum_{i\leqslant i'\leqslant j'\leqslant j}b_{j'\!j}x_{ji}\widehat b_{ii'}e_{j'\!i'}^*
=\sum_{i'\leqslant j'}(BXB^{-1})_{j'\!i'}e_{j'\!i'}^*.
\]
Therefore, the elements $\mathcal I_{ij}$, $j\leqslant i$, of the
matrix
\[
\mathcal I=BXB^{-1}, \quad B\in T(n),
\]
form a complete set of the independent lifted invariants of ${\rm
Ad}^*_{T(n)}$.
\end{proof}

\begin{note}
The center of the group $T(n)$ is $Z(T(n))=\{\beta E^n\mid\beta
\in\mathbb F/\{0\}\,\}$. If $\mathbb F=\mathbb C$ then the group
$T(n)$ is connected. In the real case the connected component
$T_{+}(n)$ of the unity in $T(n)$ is formed by the matrices from
$T(n)$ with positive diagonal elements, i.e., $T_{+}(n)\simeq
T(n)/\mathbb Z_2^n$, where $\mathbb
Z_2^n=\{\diag(\varepsilon_1,\dots,\varepsilon_n)\mid\varepsilon_i=\pm1\}$.
The inner automorphism group $\Int(\mathfrak t(n))$ of~$\mathfrak
t(n)$ is isomorphic to the factor-group $T(n)/Z(T(n))$ (or
$T_{+}(n)/Z(T(n))$ if $\mathbb F$ is real) and hence its dimension
is $\frac12n(n+1)-1$. The value of one from the diagonal elements of the matrix~$B$
or a homogenous combination of them in the above
representation of lifted invariants can be assumed inessential. It is evident
from the proof of
Theorem~\ref{TheoremOnBasisOfInvsOfCoadjRepresentationOfUpperTriangularMatrices}
that in all cases, the invariant sets of the coadjoint
representations of $\Int(\mathfrak t(n))$ and~$\mathfrak t(n)$
coincide.
\end{note}

Let us remind that $A^{i_1,i_2}_{j_1,j_2}$, where $i_1\leqslant
i_2$, $j_1\leqslant j_2$, denotes the submatrix
$(a_{ij})^{i=i_1,\ldots,i_2}_{j=j_1,\ldots,j_2}$ of a matrix
$A=(a_{ij})$. The conjugate value of $k$ with respect to $n$ is
denoted by $\varkappa$, i.e. $\varkappa=n-k+1$.

Under the proof of the below theorem the following technical lemma on matrices is used.

\begin{lemma}\label{LemmaOnEqualitiesWithSubmatrix}
Suppose $1<k<n$.
If $|X^{\varkappa+1,n}_{1,k-1}|\ne0$ then for any $\beta\in\mathbb F$
\begin{gather*}\arraycolsep=0.5ex
\beta-X^{i,i}_{1,k-1}(X^{\varkappa+1,n}_{1,k-1})^{-1}X^{\varkappa+1,n}_{j,j}=
\frac{(-1)^{k+1}}{|X^{\varkappa+1,n}_{1,k-1}|}
\left|\begin{array}{lc} X^{i,i}_{1,k-1} & \beta \\[1ex]
X^{\varkappa+1,n}_{1,k-1}& X^{\varkappa+1,n}_{j,j} \end{array}\!\right|.
\end{gather*}
In particular,
$x_{\varkappa k}-X^{\varkappa,\varkappa}_{1,k-1}(X^{\varkappa+1,n}_{1,k-1})^{-1}X^{\varkappa+1,n}_{k,k}=
(-1)^{k+1} |X^{\varkappa+1,n}_{1,k-1}|^{-1} |X^{\varkappa,n}_{1,k}|$.
Analogously
\begin{gather*}
\left(x_{\varkappa j}-X^{\varkappa,\varkappa}_{1,k-1}(X^{\varkappa+1,n}_{1,k-1})^{-1}X^{\varkappa+1,n}_{j,j}\right)
\left(x_{jk}-X^{j,j}_{1,k-1}(X^{\varkappa+1,n}_{1,k-1})^{-1}X^{\varkappa+1,n}_{k,k}\right)
\\[1.5ex]\arraycolsep=.5ex
\qquad=\frac{1}{|X^{\varkappa+1,n}_{1,k-1}|}
\left|\begin{array}{lc} X^{j,j}_{1,k} & \beta \\[1ex] X^{\varkappa,n}_{1,k}& X^{\varkappa,n}_{j,j} \end{array}\!\right|+
\frac{|X^{\varkappa,n}_{1,k}|}{|X^{\varkappa+1,n}_{1,k-1}|^2}
\left|\begin{array}{lc} X^{j,j}_{1,k-1} & \beta \\[1ex] X^{\varkappa+1,n}_{1,k-1}& X^{\varkappa+1,n}_{j,j} \end{array}\!\right|.
\end{gather*}
\end{lemma}

\begin{theorem}\label{TheoremOnBasisOfInvsOfCoadjRepresentationOfUpperTriangularMatrices}
A basis of $\Inv({\rm Ad}^*_{T(n)})$ is formed by the rational
expressions
\[\arraycolsep=.5ex
\frac{1}{|X^{\varkappa,n}_{1,k}|}\displaystyle\sum_{j=k+1}^{\varkappa-1}
\left|\begin{array}{ll} X^{j,j}_{1,k} & x_{jj} \\[1ex] X^{\varkappa,n}_{1,k}& X^{\varkappa,n}_{j,j} \end{array}\!\right|,
\quad k=0, \ldots, \left[\frac {n-1}2\right],
\]
where $|X^{n+1,n}_{1,0}|:=1$.
\end{theorem}

\begin{proof}
We choose the following normalization restriction on the lifted
invariants $\mathcal I_{ij}$, $j\leqslant i$:
\begin{gather*}
\mathcal I_{n-j+1,j}=1,\quad j=1,\ldots,\left[\frac{n}2\right],\\
\mathcal I_{ij}=0 \quad\mbox{if}\quad j\leqslant i,\quad (i,j)\not=(j',j'),(n-j'+1,j'),\quad j'=1,\ldots,\left[\frac{n+1}2\right].
\end{gather*}
This means that we do not only fix the values of the elements of the
lifted invariant matrix~$\mathcal I$, which are situated on the main
diagonal over or on the secondary diagonal. The elements of the
secondary diagonal underlying the main diagonal are given a value of~1.
The other significant elements of~$\mathcal I$ are given a value~0.
As shown below, the imposed normalization provides satisfying the conditions of
Proposition~\ref{PropositionOnNormalization} and, therefore, is correct.

Similarly to the case of strictly triangular matrices, in view of the
(triangular) structure of the matrices $B$ and $X$ the formula
$\mathcal I=BXB^{-1}$ determining the lifted invariants implies that
$BX=\mathcal IB$. This matrix equality is significant for the matrix
elements lying not over the main diagonals of the left and right
hand sides, i.e.,
\[
\sum_{i\leqslant i'}b_{ii'}x_{i'\!j}=\sum_{j'\leqslant j}\mathcal I_{ij'}b_{j'\!j}, \quad j\leqslant i.
\]
For convenience we again divide the latter system under the chosen
normalization conditions into four sets of subsystems
\begin{gather*}
S_1^k\colon\qquad \sum_{i'\geqslant\varkappa}b_{\varkappa i'}x_{i'\!j}=0, \qquad
i=\varkappa,\quad j<k,\quad k=2,\ldots,\left[\frac n2\right],
\\
S_2^k\colon\qquad \sum_{i'\geqslant\varkappa}b_{\varkappa i'}x_{i'\!j}=b_{kj}, \qquad
i=\varkappa,\quad k\leqslant j\leqslant \varkappa,\quad k=1,\ldots,\left[\frac n2\right],
\\
S_3^k\colon\qquad \sum_{i'\geqslant k}b_{ki'}x_{i'\!j}=0, \qquad
i=k,\quad j<k,\quad k=2,\ldots,\left[\frac{n+1}2\right],
\\
S_4^k\colon\qquad \sum_{i'\geqslant k}b_{ki'}x_{i'\!k}=b_{kk}\mathcal I_{kk}, \qquad
i=k,\quad j<k,\quad k=1,\ldots,\left[\frac{n+1}2\right],
\end{gather*}
and solve them one by one. The subsystem~$S_2^1$ consists of the
equations \[b_{1j}=b_{nn}x_{nj}\] which are already solved with
respect to $b_{1j}$. For any fixed $k\in\{2,\dots,[n/2]\}$ the
subsystem $S_1^k \cup S_2^k$ is a well-defined system of linear
equations with respect to $b_{\varkappa i'}$, $i'>\varkappa$, and
$b_{kj}$, $k\leqslant j\leqslant \varkappa$. We can solve the
subsystem $S_1^k$ with respect to $b_{\varkappa i'}$,
$i'>\varkappa$:
\[
B^{\varkappa,\varkappa}_{\varkappa+1,n}=-b_{\varkappa\varkappa}X^{\varkappa,\varkappa}_{1,k-1}(X^{\varkappa+1,n}_{1,k-1})^{-1},
\]
and then substitute the obtained values into the subsystem $S_2^k$.
Another way is to find the expressions for $b_{kj}$, $k\leqslant
j\leqslant \varkappa$, by the Cramer method, from the whole system
$S_1^k \cup S_2^k$ at once since only these parameters are further
considered. As a result, they have two representations via
$b_{\varkappa\varkappa}$ and~$x$'s:
\[
b_{kj}=b_{\varkappa\varkappa}\left(x_{\varkappa j}-X^{\varkappa,\varkappa}_{1,k-1}(X^{\varkappa+1,n}_{1,k-1})^{-1}X^{\varkappa+1,n}_{j,j}\right)
=\frac{(-1)^{k+1}b_{\varkappa\varkappa}}{|X^{\varkappa+1,n}_{1,k-1}|}
\left|\begin{array}{ll} X^{\varkappa,\varkappa}_{1,k-1} & x_{\varkappa j} \\[1ex]
X^{\varkappa+1,n}_{1,k-1}& X^{\varkappa+1,n}_{j,j} \end{array}\!\right|,
\]
where $k\leqslant j\leqslant \varkappa$.
In particular,
\[
b_{kk}=(-1)^{k+1}b_{\varkappa\varkappa}|X^{\varkappa+1,n}_{1,k-1}|^{-1}|X^{\varkappa,n}_{1,k}|.
\]
Analogously, for any fixed $k\in\{2,\dots,[(n+1)/2]\}$ the subsystem
$S_3^k$ is a well-defined system of linear equations with respect to
$b_{kj}$, $j>\varkappa$, and it implies
\[
B^{k,k}_{\varkappa+1,n}=-\sum_{k\leqslant j\leqslant\varkappa}b_{kj}X^{j,j}_{1,k-1}(X^{\varkappa+1,n}_{1,k-1})^{-1}.
\]
Substituting the found expressions for $b$'s into the
equations of the subsystems~$S_4^k$, we completely exclude the parameters~$b$'s and
obtain expressions of $\mathcal I_{kk}$ only via $x$'s. Thus, under
$k=1$
\[
\mathcal I_{11}=\frac1{b_{11}}\sum_{i}b_{1i}x_{i1}=\frac{b_{nn}}{b_{11}}\sum_{i}x_{ni}x_{i1}=\frac{1}{x_{n1}}\sum_{i}x_{ni}x_{i1}
=\frac{1}{x_{n1}}\sum_{i}\left|\begin{array}{ll} x_{i1} & x_{ii} \\ x_{n1}& x_{ni} \end{array}\!\right|+\sum_{i}x_{ii},
\]
where the summation range in the first sum can be bounded by 2 and
$n-1$ since for $i=1$ and $i=n$ the determinants are equal to 0. In the
case $k\in\{2,\dots,[(n+1)/2]\}$
\begin{gather*}
b_{kk}\mathcal I_{kk}=\sum_{k\leqslant i}b_{ki}x_{ik}=\sum_{k\leqslant j\leqslant\varkappa}b_{kj}x_{jk}+\!\sum_{\varkappa<i}b_{ki}x_{ik}
\\[1ex] \phantom{b_{kk}\mathcal I_{kk}}
=\sum_{k\leqslant i\leqslant\varkappa}b_{kj}\left(x_{jk}-X^{j,j}_{1,k-1}(X^{\varkappa+1,n}_{1,k-1})^{-1}X^{\varkappa+1,n}_{k,k}\right)
\\[1ex] \phantom{b_{kk}\mathcal I_{kk}}
=b_{\varkappa\varkappa}\sum_{k\leqslant i\leqslant\varkappa}
\left(x_{\varkappa j}-X^{\varkappa,\varkappa}_{1,k-1}(X^{\varkappa+1,n}_{1,k-1})^{-1}X^{\varkappa+1,n}_{j,j}\right)
\left(x_{jk}-X^{j,j}_{1,k-1}(X^{\varkappa+1,n}_{1,k-1})^{-1}X^{\varkappa+1,n}_{k,k}\right).
\end{gather*}
After using the representation for $b_{nn}$ and manipulations with
submatrices of~$X$ (see Lemma~\ref{LemmaOnEqualitiesWithSubmatrix}), we derive that
\[
\mathcal I_{kk}=
\frac{(-1)^{k+1}}{|X^{\varkappa,n}_{1,k}|}\sum_{k\leqslant i\leqslant\varkappa}
\left|\begin{array}{ll} X^{i,i}_{1,k} & x_{ii} \\[1ex] X^{\varkappa,n}_{1,k}& X^{\varkappa,n}_{i,i} \end{array}\!\right|+
\frac{(-1)^{k+1}}{|X^{\varkappa+1,n}_{1,k-1}|}\sum_{k\leqslant i\leqslant\varkappa}
\left|\begin{array}{ll} X^{i,i}_{1,k-1} & x_{ii} \\[1ex] X^{\varkappa+1,n}_{1,k-1}& X^{\varkappa+1,n}_{i,i} \end{array}\!\right|,
\]
where $k=2,\dots,[(n+1)/2]$. The summation range in the first sum
can be taken from $k+1$ and $\varkappa-1$ since for $i=k$ and
$i=\varkappa$ the determinants are equal to 0.

The combination of the found values of $\mathcal I_{kk}$ in the
following way
\[
\tilde{\mathcal I}_{00}=\sum_{j=1}^{\left[\frac{n+1}2\right]}\mathcal I_{jj}=\sum_{i}x_{ii}, \quad
\tilde{\mathcal I}_{kk}=(-1)^{k+1}\mathcal I_{kk}-\tilde{\mathcal I}_{k-1,k-1}, \quad k=1,\ldots,\left[\frac{n-1}2\right],
\]
results in the invariants $\tilde{\mathcal I}_{k'\!k'\!}$,
$k'=0,\dots,[(n-1)/2]$, from the statement of the theorem. The
functional independence of these invariants is obvious.

Under the normalization we express the non-normalized lifted invariants via $x$'s
and find a part of the parameters $b$'s of the coadjoint action via $x$'s and the other $b$'s.
No equations involving only $x$'s are obtained.
In view of Proposition~\ref{PropositionOnNormalization}, this implies
that the choice of the normalization constraints is correct, i.e.,
the number of the found functionally independent invariant is
maximal and, therefore, they form a basis of $\Inv({\rm Ad}^*_{T(n)})$.
\end{proof}

\begin{note}
An expanded form of the invariants from
Theorem~\ref{TheoremOnBasisOfInvsOfCoadjRepresentationOfUpperTriangularMatrices}
is
\[\arraycolsep=.5ex
\sum_{j=1}^n x_{jj},
\qquad
\frac{\displaystyle\sum_{j=2}^{n-1}\left|\begin{array}{ll}x_{j1}& x_{jj} \\x_{n1}& x_{nj} \end{array}\right|}{x_{n1}},
\qquad
\frac
{\displaystyle\sum_{j=3}^{n-2} \left|\begin{array}{lll}
x_{j1}& x_{j2}& x_{jj} \\
x_{n-1,1}& x_{n-1,2}& x_{n-1,j} \\
x_{n1}& x_{n2}& x_{nj}
\end{array}\right|}
{\left|\begin{array}{ll}x_{n-1,1}&x_{n-1,2}\\x_{n1}&x_{n2}\end{array}\right|},
\qquad
\ldots\ .
\]
The first invariant corresponds to the center of $\mathfrak t(n)$.
The invariant tuple ends with
\[\arraycolsep=.5ex
\frac{\Bigl|X^{\frac{n-1}2,n}_{1,\frac{n+1}2}\Bigr|}{\Bigl|X^{\frac{n+1}2,n}_{1,\frac{n-1}2}\Bigr|}
\quad \mbox{if $n$ is odd\quad and }\quad
\frac{\displaystyle\sum_{j=\frac n2}^{\frac n2+1}\left|\begin{array}{ll}
X^{j,j\vphantom{\frac n2}}_{1,\frac n2-1} & x_{jj} \\[1.3ex]
X^{\frac n2+2,n}_{1,\frac n2-1}& X^{\frac n2+2,n}_{j,j\vphantom{\frac n2}}
\end{array}\!\right|}{\Bigl| X^{\frac n2+2,n}_{1,\frac n2-1}\Bigr|}
\quad \mbox{if $n$ is even.}\]
\end{note}

\begin{corollary}\label{CorollaryOnBasisOfInvsOfUpperTriangularMatrices}
A basis of~$\Inv(\mathfrak t(n))$ consists of the rational
invariants
\[\arraycolsep=.5ex
\hat{\mathcal I}_k=
\frac{1}{|\mathcal E^{1,k}_{\varkappa,n}|}\displaystyle\sum_{j=k+1}^{n-k}
\left|\begin{array}{ll} \mathcal E^{1,k}_{j,j} & \mathcal E^{1,k}_{\varkappa,n} \\[1ex] e_{jj} & \mathcal E^{j,j}_{\varkappa,n} \end{array}\!\right|,
\quad k=0, \ldots, \left[\frac {n-1}2\right],
\]
where $\mathcal E^{i_1,i_2}_{j_1,j_2}$, $i_1\leqslant i_2$, $j_1\leqslant j_2$, denotes the matrix $(e_{ij})^{i=i_1,\ldots,i_2}_{j=j_1,\ldots,j_2}$,
$|\mathcal E^{1,0}_{n+1,n}|:=1$, $\varkappa=n-k+1$.
\end{corollary}

\begin{proof}
The symmetrization procedure for the tuple of invariants presented in
Theorem~\ref{TheoremOnBasisOfInvsOfCoadjRepresentationOfUpperTriangularMatrices}
can be assumed trivial.
To show this, we expand the determinants in each element of the tuple
and obtain, as a result, a rational expression in $x$'s.
Only the monomials from the numerator, which do not contain the `diagonal' elements $x_{jj}$,
include coordinate functions associated with noncommuting basis elements of the algebra $\mathfrak t(n)$.
More precisely, each of the monomials includes a single pair of such coordinate functions, namely,
$x_{ji'\!}$ and $x_{j'\!j}$ for some values
$i'\in\{1,\dots,k\}$, $j'\in\{\varkappa,\dots,n\}$ and $j\in\{k+1,\dots,\varkappa-1\}$.
Hence, it is sufficient to symmetrize only the corresponding pairs of basis elements.

After the symmetrization and the transposition of the matrices, we construct the following expressions
for the invariants of $\mathfrak t(n)$ associated with the invariants from
Theorem~\ref{TheoremOnBasisOfInvsOfCoadjRepresentationOfUpperTriangularMatrices}:
\[
(-1)^k\sum_{j=k+1}^{n-k}e_{jj}+\frac{1}{|\mathcal E^{1,k}_{\varkappa,n}|}\sum_{j=k+1}^{n-k}\sum_{i'=1}^k\sum_{j'=\varkappa}^n
\frac{e_{i'\!j}e_{jj'\!}+e_{jj'\!}e_{i'\!j}}2(-1)^{i'\!j'}\bigl|\mathcal E^{1,k;\hat i'}_{\varkappa,n;\hat j'}\bigr|.
\]
Here $\bigl|\mathcal E^{1,k;\hat i'}_{\varkappa,n;\hat j'}\bigr|$ denotes
the minor of the matrix $\mathcal E^{1,k}_{\varkappa,n}$ complementary to the element $e_{i'\!j'\!}$.
Since $e_{i'\!i}e_{ij'\!}=e_{ij'\!}e_{i'\!i}+e_{i'\!j'\!}$ then
\[
\sum_{i'=1}^k\sum_{j'=\varkappa}^n
\frac{e_{i'\!i}e_{ij'\!}+e_{ij'\!}e_{i'\!i}}2(-1)^{i'\!j'}\bigl|\mathcal E^{1,k;\hat i'}_{\varkappa,n;\hat j'}\bigr|=
\arraycolsep=.5ex
\left|\begin{array}{lc} \mathcal E^{1,k}_{i,i} & \mathcal E^{1,k}_{\varkappa,n} \\[1ex]
0 & \mathcal E^{i,i}_{\varkappa,n} \end{array}\!\right|
\pm\frac12|\mathcal E^{1,k}_{\varkappa,n}|,
\]
where the sign `$+$' (resp. `$-$') have to be taken if
the elements of~$\mathcal E^{1,k}_{i,i}$ are placed after (resp. before)
the elements of~$\smash{\mathcal E^{i,i}_{\varkappa,n}}$ in all the relevant monomials.
Up to constant summands,
we obtain the expressions for the elements of an invariant basis, which are adduced in the statement and
formally derived from the corresponding expressions given in
Theorem~\ref{TheoremOnBasisOfInvsOfCoadjRepresentationOfUpperTriangularMatrices}
by the replacement $x_{ij}\to e_{ji}$ and the transposition of all matrices.
That is why the symmetrization procedure can be assumed trivial in the sense described.
The transposition is necessary in order to improve the representation of invariants since $x_{ij}\sim e_{ji}$, $j\leqslant i$.
\end{proof}

\begin{note}\label{NoteOnBasisOfInvsOfUpperTriangularMatrices}
The invariants from
Corollary~\ref{CorollaryOnBasisOfInvsOfUpperTriangularMatrices}
can be rewritten as
\[
\arraycolsep=.5ex
\hat{\mathcal I}_k=
\frac{1}{|\mathcal E^{1,k}_{\varkappa,n}|}\displaystyle\sum_{j=k+1}^{n-k}
\left|\begin{array}{ll} \mathcal E^{1,k}_{j,j} & \mathcal E^{1,k}_{\varkappa,n} \\[1ex] 0 & \mathcal E^{j,j}_{\varkappa,n} \end{array}\!\right|
+(-1)^k\sum_{j=k+1}^{n-k}e_{jj},
\quad k=0, \ldots, \left[\frac {n-1}2\right].
\]
In particular, $\hat{\mathcal I}_0=\sum_j e_{jj}$.
\end{note}

\begin{note}
Let us emphasize that
a uniform order of elements from $\mathcal E^{1,k}_{i,i}$ and $\mathcal E^{i,i}_{\varkappa,n}$
has to be fixed in all the monomials under usage of the `non-symmetrized' forms of invariants
presented in Corollary~\ref{CorollaryOnBasisOfInvsOfUpperTriangularMatrices},
Note~\ref{NoteOnBasisOfInvsOfUpperTriangularMatrices} and
Theorem~\ref{TheoremOnBasisOfInvsOfSpecialUpperTriangularMatrices} (see below).
\end{note}


\section{Solvable algebra of special upper triangular matrices}\label{SectionSolvableAlgebraOfSpecialUpperTriangleMatrices}

The Lie algebra $\mathfrak{st}(n)$ of the special (i.e., having zero
traces) upper triangular $n\times n$ matrices is imbedded in a natural
way in $\mathfrak t(n)$ as an ideal. $\dim \mathfrak{st}(n)=\frac12n(n+1)-1$. Moreover,
\[
\mathfrak t(n)=\mathfrak{st}(n)\oplus Z(\mathfrak t(n)),
\]
where $Z(\mathfrak t(n))=\langle e_{11}+\dots+e_{nn}\rangle$ is the
center of $\mathfrak t(n)$, which corresponds to the one-dimensional
Abelian Lie algebra of the matrices proportional to~$E^n$. Due to
this fact we can construct a basis of $\Inv(\mathfrak{st}(n))$
without the usual calculations involved in finding the basis of
$\Inv(\mathfrak t(n))$. It is well known that if the Lie algebra
$\mathfrak g$ is decomposable into the direct sum of Lie
algebras~$\mathfrak g_1$ and~$\mathfrak g_2$ then the union of bases
of~$\Inv(\mathfrak g_1)$ and~$\Inv(\mathfrak g_2)$ is a basis
of~$\Inv(\mathfrak g)$. A basis of $\Inv(Z(\mathfrak t(n)))$
obviously consists of only one element, e.g., $e_{11}+\dots+e_{nn}$.
Therefore, the cardinality of the basis of $\Inv(\mathfrak{st}(n))$
is equal to the cardinality of the basis of $\Inv(\mathfrak {t}(n))$ minus 1, i.e., $[(n-1)/2]$.
To construct a basis of $\Inv(\mathfrak{st}(n))$, it is enough for us to rewrite
$[(n-1)/2]$ functionally independent combinations of elements from a
basis of $\Inv(\mathfrak t(n))$ via elements of $\mathfrak{st}(n)$
and to exclude the central element from the basis.

The following basis in $\mathfrak{st}(n)$ is chosen as a subalgebra
of $\mathfrak t(n)$:
\[
e_{ij}, \quad i<j, \qquad f_k=\frac{n-k}n\sum_{i=1}^k e_{ii}-\frac kn\sum_{i=k+1}^n e_{ii}, \quad  k=1,\dots, n-1.
\]
(Usage of this basis allows for the presentation of our results in
such a form that their identity with Proposition~1
from~\cite{Tremblay&Winternitz2001} becomes absolutely evident.) The
commutation relations of $\mathfrak{st}(n)$ in the chosen basis are
\begin{gather*}
[e_{ij},e_{i'\!j'\!}]=\delta_{i'\!j}e_{ij'\!}-\delta_{ij'}e_{i'\!j}, \qquad i<j,\quad i'<j'; \\
[f_k,f_{k'}]=0,  \qquad k,k'=1,\dots, n-1;\\
[f_k,e_{ij}]=0, \qquad i<j\leqslant k \quad\text{or}\quad k\leqslant i<j; \\
[f_k,e_{ij}]=e_{ij}, \qquad i\leqslant k\leqslant j,\quad  i<j
\end{gather*}
and, therefore, coincide with those of the algebra $L(n,n-1)$
from~\cite{Tremblay&Winternitz1998}, i.e.,  $L(n,n-1)$ is
isomorphic to $\mathfrak{st}(n)$. Combining this observation with
Lemma~6 of~\cite{Tremblay&Winternitz1998} results in the following
theorem.

\begin{theorem}
The Lie algebra $\mathfrak{st}(n)$ has the maximal number of
dimensions (equal to $\frac12n(n+1)-1$) among the solvable Lie
algebras which have nilradicals isomorphic to $\mathfrak t_0(n)$. It
is the unique algebra with such a property.
\end{theorem}

\begin{theorem}\label{TheoremOnBasisOfInvsOfSpecialUpperTriangularMatrices}
A basis of $\Inv(\mathfrak{st}(n))$ consists of the rational
invariants
\[\arraycolsep=.5ex
\check{\mathcal I}_k=
\frac{(-1)^{k+1}}{|\mathcal E^{1,k}_{\varkappa,n}|}\displaystyle\sum_{j=k+1}^{n-k}
\left|\begin{array}{ll} \mathcal E^{1,k}_{j,j} & \mathcal E^{1,k}_{\varkappa,n} \\[1ex] 0 & \mathcal E^{j,j}_{\varkappa,n} \end{array}\!\right|
+f_k-f_{n-k},
\quad k=1, \ldots, \left[\frac {n-1}2\right],
\]
where $\mathcal E^{i_1,i_2}_{j_1,j_2}$, $i_1\leqslant i_2$,
$j_1\leqslant j_2$, denotes the matrix
$(e_{ij})^{i=i_1,\ldots,i_2}_{j=j_1,\ldots,j_2}$, $|\mathcal
E^{1,0}_{n+1,n}|:=1$, $\varkappa=n-k+1$.
\end{theorem}
\begin{proof}
It is enough to observe (see
Note~\ref{NoteOnBasisOfInvsOfUpperTriangularMatrices}) that
\[
\check{\mathcal I}_k=(-1)^{k+1}\hat{\mathcal I}_k+\frac{n-2k}n\hat{\mathcal I}_0, \quad k=1, \ldots, \left[\frac {n-1}2\right].
\]
These combinations of elements from a basis of $\Inv(\mathfrak t(n))$
are functionally independent. They are expressed via
elements of $\mathfrak{st}(n)$. Their number is $[(n-1)/2]$.
Therefore, they form a basis of $\Inv(\mathfrak{st}(n))$.
\end{proof}

\section{Conclusion and discussion}

In this paper we extend our purely algebraic approach for computing
invariants of Lie algebras by means of moving frames
\cite{Boyko&Patera&Popovych2006, Boyko&Patera&Popovych2007} to the
classes of  Lie algebras $\mathfrak t_0(n)$, $\mathfrak t(n)$ and
$\mathfrak{st}(n)$ of strictly, non-strictly and special upper
triangular matrices of an arbitrary fixed dimension $n$. In contrast
to the conventional infinitesimal method which involves solving an
associated system of PDEs, the main steps of the applied algorithm
are the construction of the matrix~$B(\theta)$ of inner
automorphisms of the Lie algebra under consideration, and the
exclusion of the parameters~$\theta$ from the algebraic system
$\mathcal I=\check x\cdot B(\theta)$ in some way. The version of
the algorithm, applied in this paper, is distinguished in that a special usage of the
normalization procedure when the number, and a form of elements in a
functional basis of an invariant set, are determined under excluding
the parameters simultaneously.

A basis of $\Inv(\mathfrak t_0(n))$ was already known and
constructed by both the infinitesimal
method~\cite{Tremblay&Winternitz2001} and the algebraic algorithm
with an elegant but empiric technique of excluding the parameters~\cite{Boyko&Patera&Popovych2007}.
Note that the proof introduced
in~\cite{Tremblay&Winternitz2001} is very sophisticated and was
completed only due to the thorough mastery of the used infinitesimal
method. A form of elements from a functional basis of
$\Inv(\mathfrak t_0(n))$ was guessed via calculation of bases for a
number of small $n$'s and then justified with the infinitesimal
method, and both the testing steps (on invariance and on sufficiency
of number) were quite complicated.

Invariants of $\mathfrak t_0(n)$ are considered in this paper in order
to demonstrate the advantages of the normalization technique and to
pave the way for further applications of this technique to the more
complicated algebras $\mathfrak t(n)$ and  $\mathfrak{st}(n)$, being
too complex for the infinitesimal method (only the lowest few were completely
investigated there).
First the invariants of the algebras $\mathfrak t(n)$ and  $\mathfrak{st}(n)$
are exhaustively studied in this paper.
The performed calculations are simple and clear
since the normalization procedure is reduced by the choice of
natural coordinates on the inner automorphism groups and by the use
of a special normalization technique to solving a \emph{linear} system of algebraic equations.
The results obtained for $\Inv(\mathfrak{st}(n))$ in
Theorem~\ref{TheoremOnBasisOfInvsOfSpecialUpperTriangularMatrices}
completely agree with the conjecture formulated as Proposition~1
in~\cite{Tremblay&Winternitz2001} on the number and form of basis elements
of this invariant set.

A direct extension of the present investigation is to describe the
invariants of the subalgebras of~$\mathfrak{st}(n)$, which contain
$\mathfrak t_0(n)$. Such subalgebras exhaust the set of solvable Lie
algebras which can be embedded in the matrix Lie algebra
$\mathfrak{gl}(n)$ and have the nilradicals isomorphic to $\mathfrak
t_0(n)$. A~technique similar to that used in this paper can be
applied. The main difficulties will be created by breaking in
symmetry and complication of coadjoint representations. The question
on ways of investigation of the other solvable Lie algebras with the
nilradicals isomorphic to $\mathfrak t_0(n)$ remains open. (See,
e.g., \cite{Tremblay&Winternitz1998} for classification of the
algebras of such type.)

A more general problem is to circumscribe an applicability domain of
the developed algebraic method. It has been already applied only to
the low-dimensional Lie algebras and a wide range of classes of
solvable Lie algebras in~\cite{Boyko&Patera&Popovych2006,
Boyko&Patera&Popovych2007} and this paper. The next step which
should be performed is the extension of the method to classes of
unsolvable Lie algebras of arbitrary dimensions, e.g., with fixed
structures of radicals or Levi factors. An adjoining problem is the
implementation of the algorithm with symbolic calculation systems.
Similar work has already began in the framework of the general
method of moving frames, e.g., in the case of rational invariants
for rational actions of algebraic groups~\cite{Hubert&Kogan2007}.
Some other possibilities on the applications of the algorithm are
outlined in~\cite{Boyko&Patera&Popovych2007}.

\medskip

\noindent
{\bf Acknowledgments.} The work was partially supported by the National Science and Enginee\-ring
Research Council of Canada, by MITACS.
The research of R.\,P. was supported by Austrian Science Fund (FWF), Lise Meitner
project M923-N13. V.\,B. is grateful for the hospitality the Centre de
Recherches Math\'ematiques, Universit\'e de Montr\'eal.


\end{document}